\documentclass[useAMS,usenatbib]{mn2e}

\usepackage{graphicx}

\title[V4745 Sgr -- a nova above the period gap]{V4745 Sgr -- a nova above the period gap and an intermediate polar candidate}
\author[A. Dobrotka et al.]{A. Dobrotka$^{1}$\thanks{E-mail:
andrej.dobrotka@stuba.sk; retter@astro.psu.edu; asliu@bigpond.net.au}, A. Retter$^{2}$\footnotemark[1] and A. Liu$^{3}$\footnotemark[1]\\
$^{1}$Faculty of Materials Science and Technology, Slovak University of Technology in Bratislava, Paul\'{\i}nska 16, 91724 Trnava,\\ The Slovak Republic\\
$^{2}$Departement of Astronomy and Astrophysics, Penn State University, 525 Davey Lab, University Park, PA, 16802-6305, USA\\
$^{3}$Norcape Observatory, PO Box 300, Exmounth, 6707, Australia}

\begin{document}

\date{Accepted ???. Received ???; in original form ???}

\pagerange{\pageref{firstpage}--\pageref{lastpage}} \pubyear{2006}

\maketitle

\label{firstpage}

\begin{abstract}
A period analysis of CCD unfiltered photometry of V4745~Sgr (Nova Sgr 2003 \#1) performed during 23 nights in the years 2003 -- 2005 is presented. The photometric data are modulated with a period of $0.20782 \pm 0.00003~d$ ($4.98768 \pm 0.00072~h$). Following the shape of the phased light curve and the presence of the periodicity in all data sets with no apparent change in its value, we interpret this periodicity as orbital in nature and this is consistent with a cataclysmic variable above the period gap. We found a probable short-term periodicity of $0.017238 \pm 0.000037~d$ ($24.82272 \pm 0.05328~m$) which we interpret as the probable spin period of the white dwarf or the beat period between the orbital and spin period. Therefore, we propose that nova V4745~Sgr should be classified as an intermediate polar candidate, supporting the proposed link between transition-oscillation novae and intermediate polars. The mass-period relation for cataclysmic variables yields a secondary mass of about $0.52 \pm 0.05~M_{\rm \odot}$.
\end{abstract}


\section{Introduction}

Novae are a subclass of cataclysmic variable stars. In these binaries, matter from a secondary star is accreted onto the surface of a white dwarf. When this material gets sufficiently hot and dense for the white dwarf to ignite a thermonuclear runaway, it blows off the hot burning layer at the surface. This event, which leads to a rapid increase in the luminosity of the binary system, is called a nova outburst.

V4745 Sgr (Nova Sgr 2003 \#1) was discovered independently by Brown and Yamato (2003) on April 25$^{th}$ and 26$^{th}$ 2003. Optical and infrared spectroscopic confirmation was made by Ashok \& Banerjee (2003). The optical spectrum showed that it was a Fe~II nova (Williams 1992), with strong Balmer lines with P~Cyg profiles. After rapid fading the nova rebrightened again and the Balmer lines with Fe~II 4176, 4233 \AA~showed complex P~Cyg profiles (Kiss \& Jacob 2003). A detailed low and medium resolution spectroscopic study of the transition phase of the nova was presented by Cs\'{a}k et al. (2004). The spectral lines switched back to strong P Cyg profiles during the peaks of oscillations. The authors concluded that the observed repetitive rebrightening events were due to multiple episodes of mass ejection and they proposed that the transition phase in classical novae is driven by repetitive instabilities of the hydrogen shell burning on the surface of the white dwarf. They derived an absolute magnitude of the system M$_V$ = 8$^{\rm m}$.3 $\pm$ 0$^{\rm m}$.5 and the interstellar reddening to be very low ($E(B-V)<0^{\rm m}.1$) and this leads to a rough distance estimate of V4745 Sgr (between 9 and 19 kpc).

V4745~Sgr is a transient type of nova (Cs\'{a}k et al. 2004) -- a nova with oscillations during the transition phase. Several models have been suggested for the transition phase, such as stellar oscillations of the hot white dwarf, oscillations caused by the wind, formation of dust blobs that move in and out of line of sight to the nova, dwarf nova outbursts and oscillations of the common envelope (Bode \& Evans 1989, Leibowitz 1993, Warner 1995, Cs\'ak et al. 2005). Shaviv (2001) argued that super-Eddington winds can be a natural explanation for the oscillatory behaviour during the transition phase of some novae. Retter (2002) suggested another solution for the transition phase. He noticed that this phase occurs around the time when the accretion disk is re-established and proposed a possible connection between this phase and intermediate polars. Finding rapid oscilations in the light curves of transient novae is required to support this hypothesis.

Currently, there are about 50 novae with known orbital periods (Warner 2002). Typical nova periods range from 2 to 9 hours. Finding the orbital period of a nova yields an estimate of the secondary mass (Smith \& Dhillon 1998). In addition, detecting several periodicities in novae can help in classifying the system into different groups of CVs, such as magnetic systems, intermediate Polars, and/or permanent superhump systems (e.g. Diaz \& Steiner 1989, Skillman et al. 1997, 1999, Retter et al. 1997, 1998, 1999, 2002, 2003, 2005, Patterson et al. 1997, 2002, Patterson \& Warner 1998, Retter \& Leibowitz 1998, Retter \& Naylor 2000, Patterson 1999, 2001, Ak et al. 2005, Balman et al. 2005, Baptista et al. 1993, Lipkin et al. 2001, Woudt \& Warner 2001, 2002, Warner 2002, Kang et al. 2006a, 2006b). This yields valuable information about the magnetic field of the white dwarf and reveals the presence or absence of the accretion disk. The suggestion that the oscillations in the transition phase of novae are connected with intermediate polars motivated us to obtain continuous photometry of the nova V4745~Sgr and to look for short-term periods in its light curve.

In this paper we report the detection of two periodicities in our photometric data ($P_{\rm 1} = 0.20782 \pm 0.00003~d$ and $P_{\rm 2} = 0.017238 \pm 0.000037~d$). In Section 2 we present our observational material. The period analysis of the data is presented in Sec. 3 and the results are discussed in Sec. 4. 

\section{Observations}

V4745 Sgr was observed during 3 nights in 2003, 11 nights in 2004 and 9 nights in 2005. The observations included 138.8 hours and 3853 data points in total and the nightly mean length was about 6.0 hours. Table 1 presents a summary of the schedule of the observations. The photometry was carried out with a 0.3-m f/6.3 telescope coupled to a SBIG ST7E CCD camera. The telescope is located in Exmounth, Western Australia. The pixel size of the CCD is 9 x 9 microns. This camera is attached to an Optec f5 focal reducer giving an image field of view of 15 x 10 arcmin. The range of seeing for the data was 2.5 -- 3 arcsec. The exposure times were between 30 and 60 sec every 120 sec, and no filter was used. Aperture photometry was used in the reduction, with an aperture size of 12 pixels. As comparison star (check star) we used GSC 7411 1591 -- $V = 13.4$ mag (GSC 7411 2719 -- $V = 12.7$ mag) for the first night in 2003, GSC 7411 1591 -- $V = 13.4$ mag (GSC 7411 1897 -- $V = 11.9$ mag) for the rest of the 2003 data and GSC 7411 1591 -- $V = 13.4$ mag (not listed in the GSC catalog -- $V = 13.8$ mag) for the 2004 and 2005 data. The magnitudes of the stars were derived from the SBIG CCDOPS software.

\begin{figure}
\includegraphics[width=84mm]{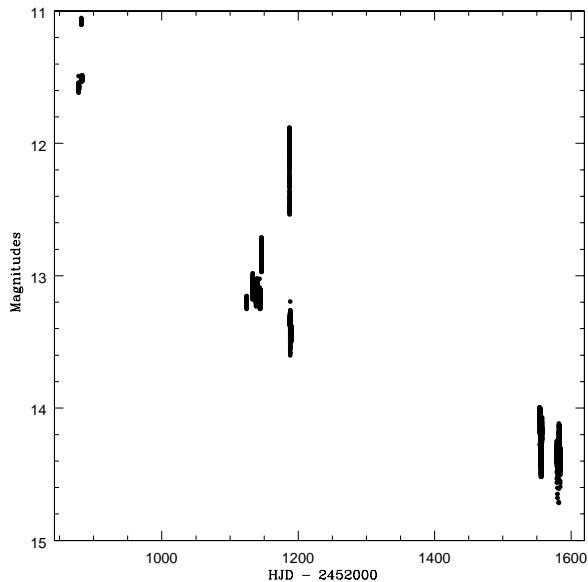}
\caption{The unfiltered light curve of our data. The $\sim 1$ mag jump observed during June 30$^{th}$ 2004 can be due to a transition oscillation of the nova.}
\label{curve}
\end{figure}

\begin{figure}
\includegraphics[width=84mm]{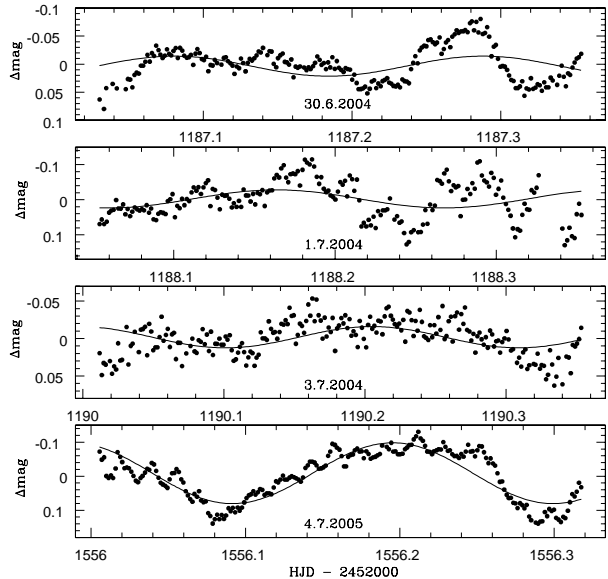}
\caption{A selection of our observations. Three detrended runs from 2004 and 1 detrended run from 2005. The solid curve is the sinusoidal fit to the data using the long-term period derived in this paper. The selected runs are discussed later in Section 3.4.}
\label{4runs}
\end{figure}

Figure~\ref{curve} displays our light curve obtained during the 3 years of coverage. The long-term light curve of the nova from the Variable Stars Network (VSNET) was published by Cs\'{a}k et al. (2004). The jump in our data at June 30$^{th}$ 2004 (HJD = 2453187) is probably due to a transition oscillation of the nova. We can not confirm this using the VSNET light curve because of the lack of observations after October 2003 in these data. In Fig.~\ref{4runs} we show 4 examples (discussed later in Section 3.4) of our observational runs with the sinusoidal fit to the data using the longer period derived in this paper. The studied variations are clearly seen.
\begin{table}
\caption{The observational log. HJD is --2450000 days, t is the duration of the observation in hours and N is number of frames.} 
\begin{center}
\begin{tabular}{rcccrcccc}
\hline
\hline
 Date & HJD (start) & t [h] & N \\
\hline
 26.08.2003 & 2877.98812 & 3.8 & 118\\
 30.08.2003 & 2882.01453 & 0.6 & 17\\
 31.08.2003 & 2882.98852 & 4.6 & 138\\
 28.04.2004 & 3124.19436 & 4.9 & 177\\
 07.05.2004 & 3133.15835 & 5.6 & 157\\
 11.05.2004 & 3137.14934 & 6.1 & 170\\
 12.05.2004 & 3138.14385 & 6.1 & 167\\
 13.05.2004 & 3139.15116 & 5.9 & 165\\
 17.05.2004 & 3143.13053 & 6.5 & 172\\
 18.05.2004 & 3144.12962 & 6.5 & 178\\
 20.05.2004 & 3146.12197 & 6.5 & 172\\
 30.06.2004 & 3187.03022 & 7.8 & 211\\
 01.07.2004 & 3188.05384 & 7.2 & 195\\
 03.07.2004 & 3190.01182 & 8.2 & 225\\
 02.07.2005 & 3554.02557 & 6.9 & 194\\
 03.07.2005 & 3555.03576 & 6.7 & 188\\
 04.07.2005 & 3556.00550 & 7.5 & 212\\
 05.07.2005 & 3557.05043 & 6.8 & 180\\
 27.07.2005 & 3579.04145 & 6.1 & 161\\
 28.07.2005 & 3580.03659 & 6.3 & 170\\
 30.07.2005 & 3582.01316 & 6.1 & 164\\
 31.07.2005 & 3583.02092 & 6.0 & 159\\
 01.08.2005 & 3584.01752 & 6.1 & 160\\
\hline
\end{tabular}
\end{center}
\end{table}

\section{Period analysis}

For the period analysis we used the Lomb-Scargle algorithm (Scargle 1982). We tested separately the 2004 and 2005 observations. The 2003 data are too few for a useful separate analysis. The data were at first detrended by substracting the linear fit from each night. Detrending by the substraction of nightly means yields similar results. A simple visual inspection of the runs suggests that the data are modulated with a periodicity of $\sim 5$ hours. The exact value of the frequency is $f_{\rm 1} = 4.812 \pm 0.007~d^{-1}$, which corresponds to the periodicity $P_{\rm 1} = 0.2078 \pm 0.0003~d$ ($4.9872 \pm 0.0072~h$). We calculated the errors in the frequency from the half width at the half maximum of the peaks in the power spectrum.

\subsection{2004 data}

\begin{figure}
\includegraphics[width=84mm]{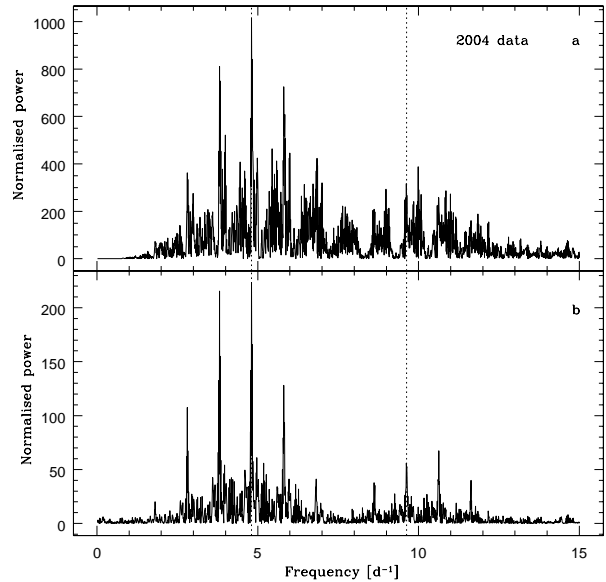}
\caption{Power spectra of the 2004 data. a -- raw data, b -- synthetic data. The dotted line at $4.812~d^{-1}$ shows the periodicity found ($f_{\rm 1}$) and at $9.624~d^{-1}$ -- its harmonic.}
\label{power_2004}
\end{figure}

The results from the period analysis of the 2004 data are depicted in Fig.~\ref{power_2004} (top panel). The suspected periodicity ($f_{\rm 1} = 4.812 \pm 0.007~d^{-1}, P_{\rm 1} = 0.2078 \pm 0.0003~d$) is clearly seen with its aliases and harmonic. To test the reality of the signal we performed a power spectrum of the synthetic data using a sinusoid with the periodicity found from the fit to the data, sampled and noised as the original data (panel b). The similarity with the original periodogram is clear. We also performed an analysis of the check star and did not identify any periodicity associated with this star.

\subsection{2005 Data}

\begin{figure}
\includegraphics[width=84mm]{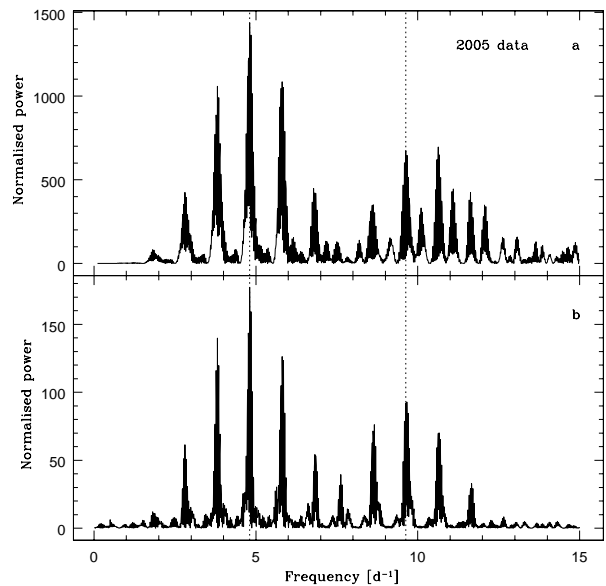}
\caption{Power spectra of the 2005 data. a -- raw data, b -- synthetic data. The dotted line at $4.812~d^{-1}$ shows the periodicity found ($f_{\rm 1}$) and at $9.624~d^{-1}$ -- its harmonic.}
\label{power_2005}
\end{figure}

The power spectrum of the 2005 light curve of V4745~Sgr is shown in the top panel of Figure~\ref{power_2005}. The strongest peak in the power spectrum ($f_{\rm 1} = 4.812 \pm 0.011~d^{-1}, P_{\rm 1} = 0.2078 \pm 0.0005~d$) is consistent with the periodicity found in 2004. To test its reliability we repeated the tests performed for the 2004 data. The synthetic power spectrum is presented in panel b. The presence of the signal and its harmonic is clear. Once again, the periodogram of the check star did not show any related periodicity.

We can then conclude that all typical features obtained by the period analysis can be explained by the periodicity $f_{\rm 1} = 4.812~d^{-1}$. The same result is obtained when we analyzed different subsamples of the data, so we conclude that this periodicity is real.

\subsection{All data}

\begin{figure}
\includegraphics[width=84mm]{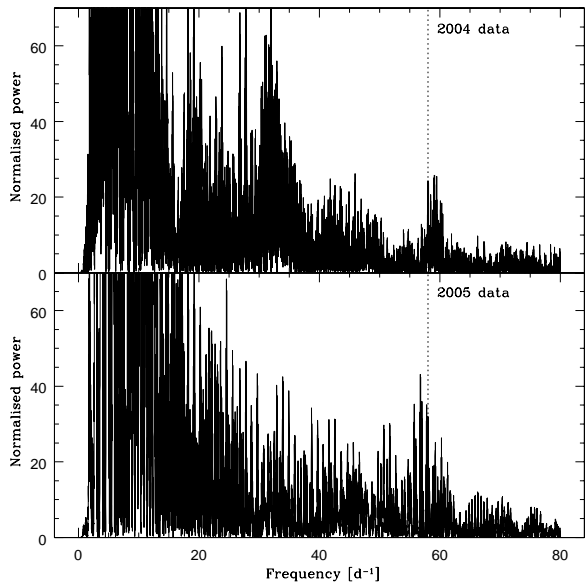}
\caption{Power spectra including the high frequency part (2004 data -- top panel, 2005 data -- bottom panel). The dashed line shows the short-term periodicity found: $f_{\rm 2} = 58.011~d^{-1}$.}
%\caption{Power spectra including the high frequency part (2004 data -- top panel, 2005 data -- bottom panel). The dashed lines show the periodicities found. From left to right: $f_{\rm 1} = 4.812~d^{-1}$, first harmonic $h_{\rm 1} = 9.6240~d^{-1}$ and short-term periodicity $f_{\rm 2} = 58.011~d^{-1}$.}
\label{power_all}
\end{figure}

The results of the period analysis up to 80 cycles/day of the 2004 and 2005 data separately are depicted in Fig.~\ref{power_all}. We analyzed each year separately. Combining the two data sets is less meaningful because of the cumulative error between the years which is larger than one full cycle (2005 -- 2004; $4.812~d^{-1} \times 0.0003~d \times 365~d = 0.5269~d > 0.2078~d$).

\subsection{Short-term oscillations}

A simple visual inspection of the runs suggests that a short-term periodicity ($\sim 0.5$ hours) is present in the data (Fig.~\ref{4runs}). In Fig.~\ref{power_all} we show the normalised power spectra of the two years separetly including the high frequency part. There is a group of peaks near the expected value, but they are very weak in power. To check whether this periodicity is real we selected different subsamples and runs and we analysed them.

\begin{figure}
\includegraphics[width=84mm]{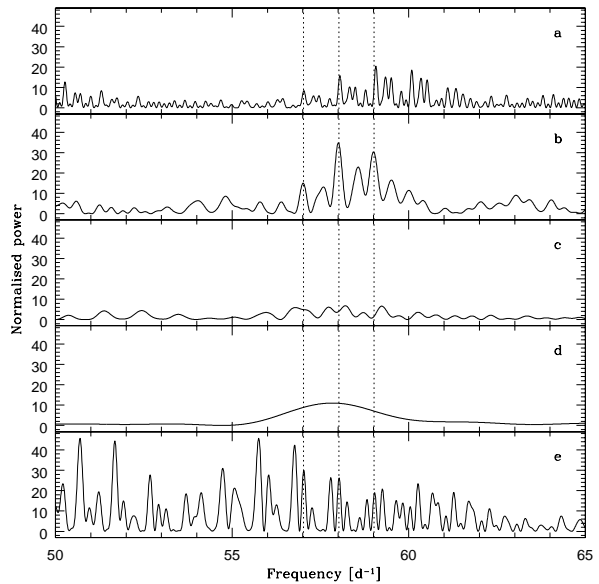}
\caption{The power spectra of different subsamples of the data arranged chronologicaly. Panel a -- nights May $13^{th}$, May $17^{th}$, May $18^{th}$ and May $20^{th}$ 2004, panel b -- June $30^{th}$, July $1^{th}$ and $3^{th}$ 2004, panel c -- July $2^{th}$, July $3^{th}$ and July $4^{th}$ 2005, panel d -- July $4^{th}$ 2005, panel e -- July $27^{th}$, July $28^{th}$, July $30^{th}$, July $31^{th}$ and August $1^{th}$ 2005. The dashed lines are from left to right: $f_{\rm 2}-1~d^{-1}$ alias, $f_{\rm 2}$ and $f_{\rm 2}+1~d^{-1}$ alias.}
\label{power_short}
\end{figure}

The first clear result was obtained using a subsample of 3 chosen runs (June $30^{th}$, July $1^{th}$ and $3^{th}$) from 2004. The power spectrum is presented in Fig.~\ref{power_short}b. We found a periodic signal of $f_{\rm 2} = 58.011 \pm 0.125~d^{-1}$ ($P_{\rm 2} = 0.017238 \pm 0.000037~d$, $0.413712 \pm 0.000888~h$). After substracting the period $P_{\rm 1}$ and its first harmonic this peak survived.

%Panel a (in both Figures) shows the power spectrum of the detrended data, panel b is the same but with fitted and extracted a sinusoid with the period $P_{\rm 1}$ and panel c is after a further extraction of the first harmonic. It is clear that the peaks are still present (short-term oscillations and aliases). Extracting only the short-term periodicity we get the power spectrum shown in panel d. The high frequency signal disappeared completely. The last panels are synthetic power spectra of sinusoids with the period $P_{\rm 1}$ and its first harmonic with the short-term periodicity (panel e) and without it (panel f).

Figure~\ref{power_short}d shows the power spectrum of one suitable night in 2005 (July $4^{th}$). We found a periodicity of $f_{\rm 2} = 57.84 \pm 1.00~d^{-1}$ which is within the errors of the previously derived value. This peak survived the test of substracting the period $P_{\rm 1}$ and its first harmonic.
%Panel a shows the periodogram of the detrended data and panel b gives the result after fitting and substracting the period $P_{\rm 1}$ and its first harmonic assuming a sinusoidal shape of the variation. The presence of the short-term signal is clear. Panel c gives the result after fitting and substracting the short-term signal which causes the disappearance of the peak and finally panel d shows the synthetic power spectrum.
It is surprising that the short-term variation is best seen only on this night in the 2005 observations. Even a simple visual inspection of the data from this night shows the prominent presence of the variation. Adding other nights increased the noise in the power spectrum and provides negative result (Fig.~\ref{power_short}c, nights -- July $2^{th}$, July $3^{th}$ and July $4^{th}$ 2005).

%The presence of the periodicity and the coincidence of the periodogram with the synthetic one are clear in both subsamples.
Other subsamples produced negatives or very unclear results (Fig.~\ref{power_short}a with dominant $f_{\rm 2}+1~d^{-1}$ alias, nights -- May $13^{th}$, May $17^{th}$, May $18^{th}$ and May $20^{th}$ 2004 or Fig.~\ref{power_short}e, nights -- July $27^{th}$, July $28^{th}$, July $30^{th}$, July $31^{th}$ and August $1^{th}$ 2005), therefore the presented cases are the best ones. The 2004 subsample is probably the best because of the night from June $30^{th}$ where the magnitude rose by $\sim 1$ mag. The signal is still weak in power, therefore our detection of the short-term periodicity needs to be confirmed by further observations. However we can consider the signal to be very probable.

Other strong peaks seen in the power spectrum are in the interval 15 -- 40 cycles per day. They are visible in Fig.~\ref{power_all} (top panel), therefore present only in the 2004 data. As there is no dominant peak we interpret these features as quasi-periodic oscillations.

\subsection{Structure of the periodicities}

\begin{figure}
\includegraphics[width=84mm]{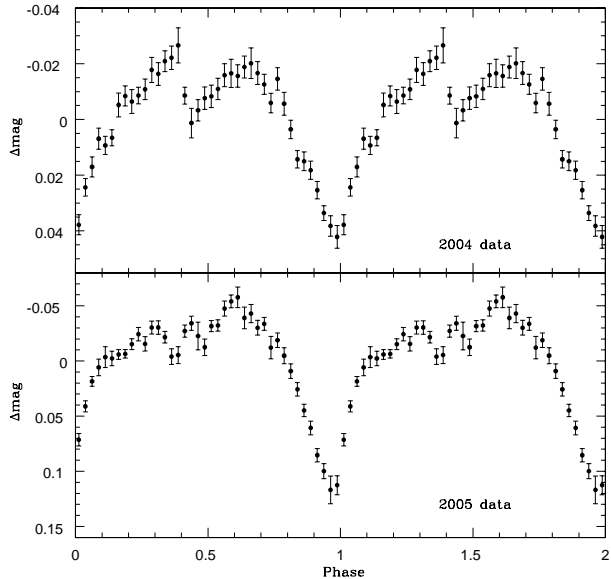}
\caption{The light curve of V4745~Sgr obtained in 2004 (top panel) and 2005 (bottom panel), folded on the $0.20782~d$ period and binned into 40 equal bins.}
\label{folded_curve_1}
\end{figure}

\begin{figure}
\includegraphics[width=84mm]{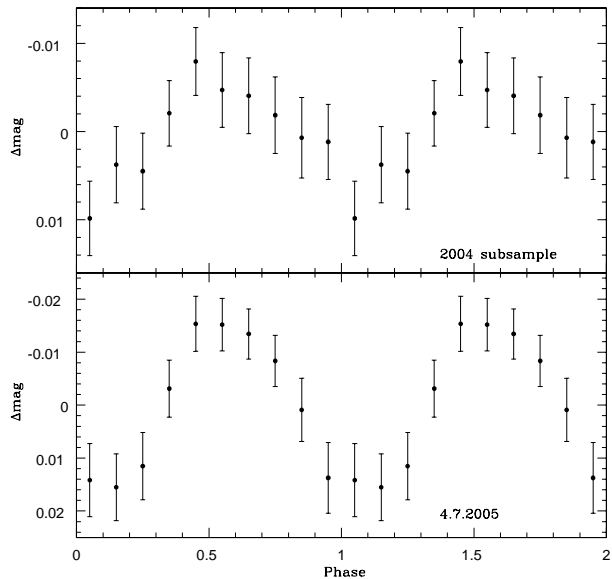}
\caption{The light curve of V4745~Sgr obtained in 2004 (top panel, subsample of the data) and 2005 (bottom panel, one night) folded on the $0.017237~d$ period and binned into 10 equal bins.}
\label{folded_curve_2}
\end{figure}

In Figure~\ref{folded_curve_1} we show the light curve of V4745~Sgr folded on the $0.20782~d$ period. The points are the average magnitude values in each of the 40 equal bins that cover the 0 -- 1 phase interval. The full amplitude of the mean variation is $0.069 \pm 0.006$ and $0.175 \pm 0.013$ mag for the 2004 and 2005 observations respectively. Using the sinusoidal fit to determine the amplitude is not adequate, because of the non sinusoidal shape of the light curve, we therefore estimated the amplitude by measuring the difference between extrema. In Figure~\ref{folded_curve_2} we present the light curve of V4745~Sgr folded on the $\sim 25$ min period and binned into 10 equal bins. The full amplitude of the mean variation is $0.013 \pm 0.001$ and $0.034 \pm 0.001$ mag for the 2004 subsample (June $30^{th}$, July $1^{th}$ and $3^{th}$) and the one night in 2005 (July $4^{th}$) respectively. The bars are the errors of the mean value. The amplitudes of the mean variations were derived by fitting a sinusoidal curve to the mean light curve.

We determided the epochs of minima by fitting low-order polynomials to the selected parts of the light curve. The best fitted (using sinusoidal fits) ephemerides of the periodicities are:
\vskip2mm
$T_{\rm 1(min)}(HJD) = 2453137.33760(53) + 0.20782(3) E$\\
\vskip2mm
$T_{\rm 2(min)}(HJD) = 2453556.20027(10) + 0.01724(3) E$\\

Since the error of the periodicity found from the periodogram analysis is larger than the error derived by the fit, we adopt instead the value $P_{\rm 1} = 0.20782 \pm 0.00003~d$ determined by the fit to the data. Since the error in the fit determination of the short-term periodicity is larger than the error from the periodogram analysis, we adopt instead the value $P_{\rm 2} = 0.017238 \pm 0.000037~d$.

The light curves folded on the $0.20782~d$ period show the typical features of an eclipse. Noteworthy is the dip at phase 0.5 with an amplitude of $\sim 0.02$ mag which is clearly present in the 2004 data. In the 2005 observations it is not so regular, but the amplitude is higher $\sim 0.05$ mag.

\section{Discussion}

We have identified two periodicities in the light curve of the nova V4745~Sgr. The first one is $P_{\rm 1} = 0.20782 \pm 0.00003~d$ ($4.98768 \pm 0.00072~h$). The period was constant during our observations and the shape of the folded light curve suggests an eclipse. We therefore propose that this periodicity is the orbital period of the binary system. Such a period is typical of orbital periods in novae (Warner 2002) above the period gap. The full amplitude of $P_{\rm 1}$ increased from 0.069 mag in 2004 to 0.175 mag in 2005. A similar behaviour was observed in three other novae; V838~Her (Leibowitz et al. 1992), V1494~Aql (Bos et al. 2002) and V4743~Sgr (Richards et al. 2005, Kang et al. 2006b). These authors interpreted the phenomenon as an eclipse of the accretion disk by the secondary star. Similarly, we propose to explain the 4.98768-hour variation in the light curve of V4745~Sgr as an eclipse of the accretion disk by the companion star.

Using the orbital period of the system, we can obtain a rough estimate for the secondary star mass. Smith \& Dhillon (1998) derived a mass-period relation for the secondary stars in CVs using a sample of reliable mass estimates. From their equation (9) we obtain a mass of $0.52 \pm 0.05~M_{\rm \odot}$ for the secondary star in V4745~Sgr. Using a mean white dwarf mass of $0.85 \pm 0.05~M_{\rm \odot}$ given for classical novae (Smith \& Dhillon 1998), we find a mass ratio of $M_{\rm 2}/M_{\rm 1} = 0.6 \pm 0.1$.

V4745~Sgr was classified as a transient type of nova (Cs\'{a}k et al. 2004). Retter (2002) proposed a possible connection between the transition nature and intermediate polars. A short-term periodicity seems to be seen in the light curve (Fig.~\ref{4runs}). We found a very probable signal of $f_{\rm 2} = 58.011 \pm 0.125~d^{-1}$ ($P_{\rm 2} = 0.017238 \pm 0.000037~d$, $0.413712 \pm 0.000888~h$). Such a signal could presumably be the spin period of the white dwarf or the beat period between the orbital period and the spin period of the white dwarf and confirm the intermediate polar nature of the system.  If the $\sim 1$ mag jump in the light curve, which was observed during June 30$^{th}$ 2004, indeed represents an oscillation during the transition phase, in the model of Cs\'ak et al. (2005), the accretion disk is destroyed. In this case, it should be easier to observe variations caused by the spin period of the exposed white dwarf. This may explain why the short-term periodicity is best seen around this selected night. We still need a confirmation for this period because the signal was only detected in a few subsamples of our light curve. Therefore we propose that nova V4745~Sgr can be a member in the intermediate polars group and this can supports the previous suggestion that the transition phase of novae may be related to intermediate polars. The best way to confirm our result is to obtain X-ray observations which are crucial in identifying of the spin period of the white dwarf. Such a confirmation was made in the case of a few other systems. No short-term periodicity was found in the optical light curve of the transient nova V1494~Aql, but a $\sim 40$ min signal was found in the X-ray data (Drake et al. 2003). A similar detection of an X-ray periodicity of $\sim 22$ min was found in the observations of the transient nova V4743~Sgr (Ness et al. 2003). Therefore the intermediate polar nature of the nova V4745~Sgr should be clarified after further X-ray and optical observations. Similar results were obtained by Ak et al. (2005). They analysed photometry of the transient nova V2540 Oph and did not detect any short-term periodicity in their data. Confirmation by X-ray observations is required in this case as well.

\section{Summary}

We found a periodic signal of 4.98768 hours in $\sim 130$ hours of unfiltered photometric observations obtained during 2 years of the nova V4745~Sgr. We interpret this periodicity as orbital and therefore the nova belongs to the numerous systems above the period gap distribution of cataclysmic variables. We found a probable short-term periodicity of $\sim 25$ minutes which needs to be confirmed. We tentatively interpret this periodicity as the spin period of the white dwarf or as the beat period between the orbital and spin period of the white dwarf. The nova should then be classified as an intermediate polar candidate.

\section*{Acknowledgments}

This work was supported by the Slovak Academy of Sciences Grant No. 4015/4.

\label{lastpage}

\end{document}